\documentclass[aic]{iosart2x}

\usepackage{comment}
\pdfoutput=1

\pubyear{0000}
\volume{0}
\firstpage{1}
\lastpage{1}

\begin{document}

\begin{frontmatter}

\title{Deep Reinforcement Learning for Multi-Agent Interaction}
\runtitle{Deep Reinforcement Learning for Multi-Agent Interaction}



\begin{aug}
    \author{\inits{I.H.}\fnms{Ibrahim H.} \snm{Ahmed}\ead[label=e1]{i.ahmed@ed.ac.uk}}%
    \author{\inits{C.B.}\fnms{Cillian} \snm{Brewitt}\ead[label=e2]{cillian.brewitt@ed.ac.uk}}
    \author{\inits{I.C.}\fnms{Ignacio} \snm{Carlucho}\ead[label=e3]{ignacio.carlucho@ed.ac.uk}}
    \author{\inits{F.C.}\fnms{Filippos} \snm{Christianos}\ead[label=e4]{f.christianos@ed.ac.uk}}
    \author{\inits{M.D.}\fnms{Mhairi} \snm{Dunion}\ead[label=e5]{mhairi.dunion@ed.ac.uk}}
    \author{\inits{E.F.}\fnms{Elliot} \snm{Fosong}\ead[label=e6]{e.fosong@ed.ac.uk}}
    \author{\inits{S.G.}\fnms{Samuel} \snm{Garcin}\ead[label=e7]{s.garcin@ed.ac.uk}}
    \author{\inits{S.G.}\fnms{Shangmin} \snm{Guo}\ead[label=e8]{s.guo@ed.ac.uk}}
    \author{\inits{B.G.}\fnms{Balint} \snm{Gyevnar}\ead[label=e9]{balint.gyevnar@ed.ac.uk}}
    \author{\inits{T.M.}\fnms{Trevor} \snm{McInroe}\ead[label=e10]{t.a.mcinroe@sms.ed.ac.uk}}
    \author{\inits{G.P.}\fnms{Georgios} \snm{Papoudakis}\ead[label=e11]{g.papoudakis@ed.ac.uk}}
    \author{\inits{A.R.}\fnms{Arrasy} \snm{Rahman}\ead[label=e12]{arrasy.rahman@ed.ac.uk}}%
    \author{\inits{L.S.}\fnms{Lukas} \snm{Schäfer}\ead[label=e13]{l.schaefer@ed.ac.uk}}
    \author{\inits{M.T.}\fnms{Massimiliano} \snm{Tamborski}\ead[label=e14]{m.tamborski@sms.ed.ac.uk}}
    \author{\inits{G.V.}\fnms{Giuseppe} \snm{Vecchio}\ead[label=e15]{g.vecchio@sms.ed.ac.uk}}
    \author{\inits{C.W.}\fnms{Cheng} \snm{Wang}\ead[label=e16]{cheng.wang@ed.ac.uk}}
    \author{\inits{S.A.}\fnms{Stefano V.} \snm{Albrecht}\ead[label=e17]{s.albrecht@ed.ac.uk}
    \thanks{Corresponding author. \printead{e17}.}}
    \address{Autonomous Agents Research Group, School of Informatics, \orgname{University of Edinburgh}, \cny{United Kingdom} 
    }
\end{aug}

\begin{abstract}
    The development of autonomous agents which can interact with other agents to accomplish a given task is a core area of research in artificial intelligence and machine learning. Towards this goal, the Autonomous Agents Research Group develops novel machine learning algorithms for autonomous systems control, with a specific focus on deep reinforcement learning and multi-agent reinforcement learning. Research problems include scalable learning of coordinated agent policies and inter-agent communication; reasoning about the behaviours, goals, and composition of other agents from limited observations; and sample-efficient learning based on intrinsic motivation, curriculum learning, causal inference, and representation learning. This article provides a broad overview of the ongoing research portfolio of the group and discusses open problems for future directions.
\end{abstract}


\end{frontmatter}

\section{Autonomous Agents Research Group}

The {\it Autonomous Agents Research Group}\footnote{Homepage: \url{https://agents.inf.ed.ac.uk}, \ Blog: \url{https://agents.inf.ed.ac.uk/blog}} is a research group led by Dr. Stefano V. Albrecht in the School of Informatics, University of Edinburgh. The long-term goal of the group is to develop artificial intelligence and machine learning technologies which enable autonomous agents (such as robots and software agents) to solve tasks in complex environments. The group has a strong focus on problems of coordination and cooperation in multi-agent systems, in which multiple autonomous agents interact in a shared environment. Current research focuses on algorithms for deep reinforcement learning (RL) and multi-agent reinforcement learning (MARL). The group is also involved in the development of industry applications, including in the areas of autonomous driving (with industry partner Five AI) and multi-robot warehouse logistics (with industry partner Dematic/KION). We are a member of the ELLIS European network of excellence in machine learning research.

This article provides an overview of the work conducted in our group and our research contributions. We start by providing a brief overview of the main research strands in Section~\ref{sec:summary}, followed by more detailed descriptions of research highlights in Section~\ref{sec:highlights}. Section~\ref{sec:codebase} provides descriptions of several of our open-source code repositories. Finally, in Section~\ref{sec:openproblems} we discuss important open problems in the literature before concluding in Section~\ref{sec:conclusion}.


\section{Research Strands}
\label{sec:summary}
 
Research in the group focuses on the following research strands:

\begin{description}

    \item[Multi-Agent Reinforcement Learning]
    Scalable learning of coordinated agent policies and inter-agent communication in multi-agent systems is a long-standing open problem. We tackle this problem by developing algorithms for multi-agent deep RL, in which multiple agents learn how to communicate and (inter-)act optimally to achieve a specified goal \cite{papoudakis2019dealing,papoudakis2021benchmarking,christianos2020shared,christianos2021scaling,guo2022expressivity,schaefer2022mate}. While deep RL has enabled scalability to large state spaces, the goal of MARL is to allow efficient scalability in the number of agents where the joint decision space would otherwise be intractable for centralised approaches. \\
    
    \item[Decision Making and Modelling Other Agents]
    Our long-term goal is to create autonomous agents capable of robust goal-directed interaction with other agents, focusing on ad hoc teamwork problems \cite{mirsky2022survey,albrecht2017special,rahman2021open,papoudakis2021local} that require fast and effective adaptation without opportunities for prior coordination between agents. We develop algorithms which enable agents to reason about the behaviours, capabilities, and composition of other agents from limited observations \cite{albrecht2018modelling,albrecht2020special}. These inferences are used in combination with RL and planning techniques for effective decision-making. \\
    
    \item[Single-Agent Deep Reinforcement Learning]
    While the group's primary focus is on multi-agent systems, 
    we are also actively contributing to research in areas of single-agent RL. 
    We develop single-agent RL algorithms that can learn optimal and robust policies through minimal interactions with the environment. To this end, we are developing algorithms that leverage techniques such as intrinsic motivation, curriculum learning, causal inference, and representation learning in high-dimensional state spaces~\cite{mcinroe2021learning,mcinroe2022learning,dunion2022ted,schaefer2022derl} to achieve sample-efficient learning. \\ 

    \item[Autonomous Driving in Urban Environments]
    We develop algorithms for autonomous driving in challenging urban environments, enabling autonomous vehicles to make fast, robust, and safe decisions by reasoning about the actions and intent of other actors in the environment~\cite{albrecht2020igp2,hanna2021interpretable,brewitt2021grit,brewitt2022verifiable,gyevnar2022humancentric,vecchio2022midgard}. Research topics include: complex state estimation in uncertain and dynamic environments; efficient reasoning about intent from limited data; and computing robust plans with specified safety-compliance under conditions of dynamic, uncertain observations and limited compute budget. We collaborate closely with Five AI, a UK-based company developing autonomous driving technologies. \\

    \item[Quantum-Secure Authentication and Key Agreement]
    Classical protocols for authentication and key establishment relying on public-key cryptography are vulnerable to quantum computing. We develop a novel, quantum-resistant approach to authentication and key agreement based on the complexity of interaction in multi-agent systems, supporting mutual and group authentication and forward secrecy~\cite{ahmed2021quantum}. We leverage recent progress in generative adversarial training \cite{wiatrak2019stabilizing} and deep RL to maximise our system's security against intruders and modelling attacks.
\end{description}

\section{Research Highlights}
\label{sec:highlights}    

    \subsection{Multi-Agent Reinforcement Learning}

    One issue that predominantly affects MARL research is the lack of reproducibility of state-of-the-art algorithms and the insufficient number of standardised benchmark environments. As a result, measuring the progress of research in MARL proves challenging. To address this issue, we benchmarked nine MARL algorithms in five multi-agent environments within which we defined 25 different cooperative tasks~\citep{papoudakis2021benchmarking}. We evaluated three classes of MARL algorithms: independent learners, centralised policy gradient learners, and value decomposition learners. Additionally, we open-sourced two multi-agent environments, Level-Based Foraging (LBF) and the Multi-Robot Warehouse (RWARE), as well as the \textit{Extended} PyMARL (EPyMARL) repository, which extends PyMARL~\citep{samvelyan19smac} to include more algorithms, additional implementation details, and compatibility with more environments. We provide more information about the environments and open-source repositories in Section~\ref{sec:codebase}. Our research~\citep{papoudakis2021benchmarking} presented standardised performance metrics throughout training and provided the intuition behind our evaluation findings.
    
    Exploration is a challenging problem in MARL, especially in sparse reward settings. Our work, \textit{Shared Experience Actor} (SEAC)~\cite{christianos2020shared}, approached this problem by combining the experiences generated by different agents into more informative learning gradients using the basic idea of off-policy correction via importance weighting. In this work, we also discussed how the distinct behaviours of SEAC agents help with exploration. Our results showed that SEAC significantly improves sample efficiency and greatly increases returns at policy convergence.

    MARL systems with a large number of agents can lead to inefficient training of agent policies. Parameter sharing between agents is a common approach in MARL to improve training efficiency since it decreases the number of trainable parameters. However, na\"ively sharing all parameters between all agents leads to a lack of diversity in agent behaviour. To avoid this, in our work \textit{Selective Parameter Sharing} (SePS)~\citep{christianos2021scaling}, we developed a method to automatically identify agents that can benefit from parameter sharing. SePS uses an encoder-decoder architecture to encode agent identities into an embedding space. Then, depending on the reward and observation functions of the individual agents, identity embeddings are clustered with an unsupervised clustering method. Furthermore, we provided a rigorous empirical analysis of the impact of parameter sharing and showed that SePS can be coupled with existing MARL algorithms to achieve significantly improved performance.
    
    Inter-agent communication is a common method for facilitating cooperation among agents. Such cooperation leads to agents negotiating a communication protocol during the learning process~\cite{lazaridou2016multi}. In a discrete communication channel, such emergent communication protocols are referred to as ``emergent languages" since they use discrete symbols~\cite{mul2019mastering}. Similar to how natural languages help humans complete tasks, we expect emergent languages to assist agents in completing their given tasks. The degree to which agents can embed relevant knowledge in their communication channels is referred to as the ``expressivity" of their communication. In our work~\cite{guo2022expressivity}, we demonstrated that mutual information is an insufficient measurement for expressivity. Instead, we proposed to measure expressivity as a partial ordering of languages, ranked by their generalisation across tasks. Ultimately, we found that expressivity of emergent languages is a trade-off between the complexity and unpredictability of the context for interpreting the messages sent from speaking agents to acting agents. As a result, we established that we need to increase the complexity and variation of samples in batches during training to achieve communication protocols that are universally helpful across different tasks.

\subsection{Ad Hoc Teamwork}
\label{sec:ad_hoc_teamwork}

    Real-world multi-agent applications may include robotic agents from different manufacturers or robots with heterogeneous physical characteristics and abilities. However, jointly training such groups of agents may not be feasible or possible for various reasons. Therefore, we have focused on the problem known as ad hoc teamwork, wherein agents need to be capable of cooperating on the fly without prior coordination.
    The ad hoc teamwork challenge is rooted in real-world robotics applications and was described in the seminal paper by Stone et al. \cite{stoneAdHocAutonomous2010}. Since the introduction of the challenge, a substantial number of publications have addressed different aspects of the ad hoc teamwork problem \cite{mirsky2021penny,melo2016ad}.

    In many multi-agent applications, an autonomous agent must be able to cooperate on the fly with diverse types of other agents that may dynamically enter and leave the environment. This problem is known as \textit{open} ad hoc teamwork and is a significant problem faced in the ad hoc teamwork setting. To address team openness, we introduced \textit{Graph-based Policy Learning} (GPL) \cite{rahman2021open}. Our method leverages graph neural networks (GNNs) to train a policy that is robust to dynamic team composition and team sizes. GPL learns a joint action value model based on coordination graphs, allowing the learning agent to discern the effects of each teammate's action on the overall return. Furthermore, we enhanced the joint action value computation with an action prediction module that learns to predict teammates' actions. By modelling the uncertainty in teammates' actions selection, we improved the learner's decision-making capabilities. 
    Standard RL algorithms can then use the learned action values to obtain optimal policies in the ad hoc teamwork setting. We tested GPL in several multi-agent environments and compared it against a number of baselines, showing that our method can generalise better in open team settings and under previously unseen team compositions. 
    
    An important ability to solve ad hoc teamwork problems is agent modelling, wherein an agent reasons about the behaviour of other agents. However, agent modelling under the assumption of partial observability is an open problem~\cite{albrecht2018modelling}. To address this issue, we proposed \textit{Local Information Agent Modelling} (LIAM)~\cite{papoudakis2021local}. LIAM is an encoder-decoder model that learns representations that connect the trajectory of the controlled agent with the trajectory of the modelled agent. LIAM is trained in a centralised fashion by using the trajectories of all agents in the environment to learn representations of different agent policies. However, during execution, the controlled agent operates in a decentralised manner by generating representations of the modelled agent using only the local trajectory of the controlled agent. 
    We evaluated LIAM in several multi-agent environments, and demonstrated that LIAM achieves robust agent modelling under significant partial observability.
    
    Our work in ad hoc teamwork extends beyond the development of novel algorithms. 
    Recently, we organised a workshop on ad hoc teamwork\footnote{\url{https://sites.google.com/view/ad-hoc-teamwork}} held at the International Joint Conference on Artificial Intelligence (IJCAI) 2022, in which we discussed recent advances in the field. 
    Additionally, we conducted a survey on ad hoc teamwork~\cite{mirsky2022survey} with three clear objectives in mind. First, we provided a clear definition of the scope of the ad hoc teamwork problem by outlining the key assumptions and subtasks required to address the problem in full.
	Second, we provided a concise description of current ad hoc teamwork literature regarding the types of solutions used and the evaluation domains of these solutions. Third, we discussed current open problems in the ad hoc teamwork literature.

    \subsection{Single-Agent Reinforcement Learning}
    
    While research presented so far addresses the problem of coordination in systems with multiple agents, developing a single agent that can perform complex tasks autonomously in real-world scenarios remains still largely unresolved. In such scenarios, we require agents that are able to learn policies with minimal interactions with the environment, while at the same time achieving high generalisation capabilities. To achieve this goal, we investigate several topics such as intrinsically-motivated exploration, policy evaluation, and casual models. In the following paragraphs, we highlight our published work in single-agent RL, and briefly explain our contributions and the results found.

    Intrinsic rewards are an effective method for encouraging RL agents to explore~\citep{barto2013intrinsic,oudeyer2009intrinsic}. However, the exploration process may suffer from instability caused by non-stationary reward shaping and sensitivity to hyperparameters. To address the instabilities arising from these challenges, we proposed \textit{Decoupled RL} (DeRL)~\cite{schaefer2022derl}, a general framework that trains separate policies for intrinsically-motivated exploration and exploitation. 
    DeRL trains an exploration policy on the combined intrinsic and extrinsic rewards; the exploration policy is used to collect data for the exploitation policy, which is trained only on extrinsic rewards. DeRL's training regime decouples the exploitation policy from the instability introduced by intrinsic rewards while making use of the exploration of these intrinsic rewards. We evaluated DeRL algorithms in two sparse-reward environments with multiple types of intrinsic rewards. Our results showed that intrinsically-motivated RL suffers from instability while DeRL is more robust to varying scales and rate of decay of intrinsic rewards. We also found that DeRL converges to the same evaluation returns in fewer interactions as intrinsically-motivated baselines in several tasks. Lastly, we discussed the challenge of distribution shift introduced by diverging exploration and exploitation policies throughout training. Such divergence can introduce instability in itself, and we showed that divergence constraint regularisers could successfully minimise such instability caused by the divergence of both trained policies.
    
    Policy evaluation deals with estimating the expected returns of a given policy in an environment of interest. Many works have studied how to efficiently use a static set of previously collected data for policy evaluation~\cite{precup2000eligibility,thomas2016data-efficient}. However, comparatively less research has considered how to improve data collection for data-efficient policy evaluation. One widely-used method is to use the evaluation policy to collect a set of i.i.d.\ trajectories and then use the Monte Carlo estimator that computes the average return as its estimate. Given an infinite number of trajectories, the likelihood of each trajectory in the collected set will converge to its true probability under the evaluation policy with Monte Carlo return estimates converging to the true returns under the evaluation policy. However, given a finite number of samples, such i.i.d.\ data collection may yield high variance estimates due to sampling error caused by observing trajectories in the data at a different proportion than their true likelihood under the evaluation policy. Our work showed how non-i.i.d.\ sampling can lower sampling error and increase the accuracy of the Monte Carlo estimator on a (finite) collected set of trajectories~\cite{zhong2021robust}. We introduced two non-i.i.d.\ data collection methods for policy evaluation, both of which consider previously collected data when collecting future data to reduce sampling error in the entire data set without using off-policy corrections. Both data collection methods follow the simple idea that we should more frequently sample trajectories that are underrepresented in our data with respect to their true likelihood under the evaluation policy. Equally, trajectories that occur at higher proportions in the data than their true likelihood under the evaluation policy should be sampled less frequently in further data collection. Our empirical results showed that our methods produce data with lower sampling error for finite-sized data sets and lead to lower mean-squared error in policy evaluation for any data set size compared to i.i.d.\ sampling.
    
    \subsection{Autonomous Driving}

    We have partnered with UK-based company Five AI to develop planning and prediction methods for autonomous driving, with a particular focus on goal recognition, interpretability, and verifiability.

We developed an integrated goal recognition and motion planning system, called \textit{Interpretable Goal-based Prediction and Planning} (IGP2)~\cite{albrecht2020igp2}. IGP2 infers posterior probabilities of possible goals and trajectories of other vehicles using rational inverse planning. The goal and trajectory predictions inform a motion planner based on Monte Carlo Tree Search (MCTS), which repeatedly forward-simulates the present states to determine the best sequence of actions for the ego vehicle. By modelling the actions of all vehicles using high-level manoeuvres and macro actions, IGP2 generates manoeuvre plans over extended horizons for which we can extract intuitive explanations based on rationality principles. We tested IGP2 in diverse simulated urban driving scenarios, demonstrating its ability to robustly discover goals and driving intentions of nearby vehicles and exploit this information to generate efficient driving plans for the ego vehicle.

We extended IGP2 by adding the ability to infer the presence of occluded objects. Our approach, \textit{Goal and Occluded Factor Inference} (GOFI)~\cite{hanna2021interpretable}, jointly models the probability of occluded objects and the goals of other vehicles. If an observed vehicle's behaviour seems rational given the presence of an occluded object but seems irrational with no occluded object present, then GOFI uses this information to increase its belief that an occluded object is present. Like IGP2, GOFI uses the inferred goal and occluded object probabilities to inform an MCTS planning procedure. In a range of simulated driving scenarios, we showed that GOFI is able to reduce the number of collisions with occluded objects or other vehicles.
    
Finally, we developed a goal recognition method for autonomous vehicles called \textit{Goal Recognition with Interpretable Trees} (GRIT)~\cite{brewitt2021grit}. GRIT can infer the goals of vehicles by constructing decision trees from vehicle-trajectory data. For goal recognition methods to be useful and safe in real-world settings, they must be fast, accurate, interpretable, and verifiable. Unlike previous methods, GRIT is the first method which satisfies all four of these objectives. We evaluated GRIT across four urban driving scenarios from two vehicle-trajectory datasets. Our experiments found that GRIT achieved similar accuracy to deep learning based methods and higher accuracy than several other baselines. We also showed that GRIT produces trees which are human-interpretable and that can be formally verified by mapping them into propositional logic and applying satisfiability modulo theories (SMT) solvers.

    \subsection{Secure Authentication and Key Agreement}

    Secure authentication and key agreement (AKE) protocols are the foundation for communication over computer networks. Existing protocols, whether symmetric or asymmetric in design, rely on number-theoretic problems for their security and are vulnerable to recent advances in quantum computing. As computing power increases at a fast pace, cryptographic security based on computational hardness cannot be persistently guaranteed from a mathematical perspective. To counteract this issue, we introduced a symmetric AKE protocol called \textit{Authentication via Multi-Agent Interaction} (AMI)~\cite{ahmed2021quantum}. AMI is aligned with information-theoretic security, which does not rely on computational hardness and is quantum-safe. 
    AMI is a novel formulation of symmetric AKE as a multi-agent system, where communicating parties are treated as autonomous agents whose behaviour within the protocol is governed by private agent models used as the long-term master keys. AMI's multi-agent interaction process produces interaction transcripts used for authentication, generation of session keys, and a key-evolving scheme for forward secrecy. 
    
    In~\citep{ahmed2021quantum}, we provided an authentication test based on a statistical hypothesis testing algorithm \cite{albrecht2015criticising}, which we showed to be highly accurate in not only recognising legitimate parties, but also in detecting different adversarial strategies utilising data observed from prior protocol sessions. These include a random attack, a replay attack, and a key recovery attack (model reconstruction via maximum likelihood estimate). Outside of~\cite{ahmed2021quantum}, we also developed authentication tests based on a neural network classifier, trained on interaction transcripts in two different ways - via supervised learning, and within a generative adversarial network (GAN) \citep{wiatrak2019stabilizing}. In particular, the use of a GAN allows for generative modelling of a legitimate system user, as a unique adversarial training mechanism against imitation attacks. The benefit of classifier-based authentication is that the master key is not required at run-time, which is a significant departure from existing symmetric AKE protocols.

    AMI can also be extended beyond a simple client-server setting for group authentication, wherein a cluster of more than two agents in a larger multi-agent system establish secure communication for either a centralised system (trusted third-party is present) or decentralised system. We released the PyAMI open-source framework supporting different modes of one-way, mutual, and group authentication, to practically demonstrate the AMI protocol on a live network. PyAMI consists of a multi-agent system where agents run on geographically distant remote machines and communicate over network sockets using TCP. During an interaction process, server and client machines transmit actions over the network to build a shared interaction transcript. After successful authentication, the remote parties compute an identical session key using the AMI key agreement algorithm.
    
    Finally, we demonstrated how agent models can be optimised to achieve desirable behaviour within the protocol. More specifically, we used the PPO reinforcement learning algorithm \cite{schulman2017ppo} to train the server's behavioural model (as a neural network) for sample-efficient authentication, wherein the server intelligently probes the opposing party with fewer but higher-quality queries. The training objective was to produce an interaction transcript between server and adversarial client which resulted in a sufficiently low $p$-value in the statistical hypothesis test, from the fewest number of time steps possible. We showed that the number of required samples to reject an adversary can thus be decreased by up to 70\%. An additional benefit of having fewer interactions is that less data from a private master key is publicly observable, which can strengthen a protocol's information-theoretic security. 
    
\section{Code Repository}
\label{sec:codebase}
Our group maintains a substantial collection of open-source code repositories\footnote{ \url{https://github.com/uoe-agents}}. These repositories provide implementations of all of our published algorithms and are actively maintained. In this section, we provide a brief description of some of our main repositories. 

One of our most prominent open-source repositories is EPyMARL\footnote{\url{https://github.com/uoe-agents/epymarl}}~\cite{papoudakis2021benchmarking}. EPyMARL is based on PyMARL~\cite{samvelyan19smac}, a MARL framework that includes the implementations of basic MARL algorithms. 
EPyMARL extends PyMARL by including additional algorithms, such as IA2C~\cite{Chu2020IA2C}, IPPO~\cite{schulman2017ppo}, MADDPG~\cite{LoweMADDPG}, 
and MAPPO~\cite{Yu2021TheSE}. 
Our implementation focuses on consistency, which allows for fair comparisons between algorithms and provides additional implementation options that give users flexibility. For example, one of these options enables the user to select between sharing and no-sharing of parameters between agents. Furthermore, EPyMARL includes options to choose implementation details such as hard or soft target-network updates, entropy regularisation, reward standardisation, and fully-connected or recurrent networks. In contrast to PyMARL, which is limited to training algorithms in the Starcraft Multi-Agent Challenge (SMAC)~\cite{samvelyan19smac}, EPyMARL provides general support for multi-agent environments following the standard Gym API.

The open-source implementation of our integrated goal recognition and motion planning system IGP2~\cite{albrecht2020igp2} for autonomous vehicles is available in the repository called IGP2\footnote{ \url{https://github.com/uoe-agents/IGP2}}.
The repository provides instructions on how to install and run IGP2 and is accompanied by a blog post that further explains our algorithm with videos and images. The IGP2 goal recognition module can be run in two different ways: i) using state-of-the-art datasets, such as inD~\cite{inDdataset} and roundD~\cite{rounDdataset} or ii) using the CARLA simulator~\cite{Dosovitskiy17}. Additionally, we include a visualisation tool that can be used in conjunction with the datasets to provide information in real-time.

We provide an implementation of PyAMI~\citep{ahmed2021quantum} code and documentation\footnote{\url{https://github.com/uoe-agents/PyAMI}}. PyAMI is a Python implementation of AMI, the authentication and key generation protocol developed in~\citep{ahmed2021quantum}. It runs the AMI protocol in a multi-agent system consisting of multiple remote machines communicating over network sockets via TCP (it may also be run entirely on the local machine in simulation mode for ease of use). Instructions are provided for the local directory structure, as well as the JSON file format for agent parameters on each remote machine. Run commands are also provided for supported deployments, as either a centralised or decentralised multi-agent system design.

Additionally, our repository hosts three of our custom multi-agent environments. 
First, Level-Based Foraging (LBF)\footnote{\url{https://github.com/uoe-agents/lb-foraging}}~\cite{albrecht2013game} is a set of mixed cooperative-competitive tasks that require coordination. Agents in LBF move around a grid-world map and collect items. Both items and agents are assigned a level such that agents can only pick up an item if the sum of the levels of the agents involved is equal to or greater than the item level. 
%
The second environment is the Multi-Robot Warehouse (RWARE)\footnote{\url{https://github.com/uoe-agents/robotic-warehouse}}, which simulates a warehouse with robots moving and delivering requested goods. RWARE is challenging due to sparse rewards, collision dynamics, and multi-step tasks.
The repository includes variants of the environment that allow the user to change the size of the map, the number of agents, and the difficulty level. 
The third environment is PressurePlate\footnote{\url{https://github.com/uoe-agents/pressureplate}}. In this environment, a group of agents must cooperate in traveling from one end of a gridworld to the other. Passage between rooms is blocked by doors that only open when an agent remains standing on a plate on the ground. PressurePlate is difficult because plates only respond to specific agents, and completing the task requires some agents to stay behind while the remainder of the group progresses through the environment.
These environments are built in Python and use the standard OpenAI Gym API format. For further details, see our blog post on MARL environments\footnote{\url{https://agents.inf.ed.ac.uk/blog/multiagent-learning-environments/}}.

\section{Open Problems}
\label{sec:openproblems}

In the following subsections, we outline current open problems in the literature as prioritised by our own research group. We believe finding solutions to these problems will contribute towards achieving autonomous agents that can be robustly deployed in practical applications. 

\subsection{Generalisation in Reinforcement Learning}

A key factor preventing the broader adoption of RL algorithms is its comparatively limited ability to generalise to even small variations from its training conditions. In comparison, generalisation in supervised learning settings is well understood~\cite{kawaguchi2017generalization}. Despite recent efforts to characterise and define generalisation in the context of RL~\cite{kirk2021survey,malik2021generalizable,ghosh2021generalization}, a more precise formalisation is still required.
Indeed, generalisation is a complex problem within RL and includes many different challenges. Each generalisation problem is primarily determined by the similarities and differences across tasks that agents are trained and evaluated within. Without any assumptions, training and evaluation tasks could be arbitrarily different and hence no generalisation could be feasibly expected. Fruitful research in RL generalisation requires that we enforce consistency in task aspects such as state and action spaces, transition dynamics, or reward functions. Moving from single-agent to multi-agent systems further increases the complexity and number of factors impacting the challenge of generalisation~\citep{schaefer2022mate}. In multi-agent systems, tasks might include a varying number of agents, and agents also need to generalise to other agents with which they will cooperate or compete. In particular, the latter challenge directly relates to ad hoc teamwork discussed in Section \ref{sec:ad_hoc_teamwork}.

One commonly employed approach to address generalisation in RL is to train the agent over multiple tasks that are randomly sampled from a distribution~\cite{DR}. The definition of this distribution characterises the type of generalisation being targeted. However, this process is not sample efficient. Indeed, randomly sampled tasks may be too easy or too hard, given the agent's current stage of training. On the other hand, if the tasks are not sampled but randomly generated, they may be irrelevant to the target task distribution or even be impossible to solve. 

For generalisation in the supervised learning setting, the training data distribution can be as important as the choice of the learning algorithm. Some recent research has also shown that this is the case in the RL setting. Curating training tasks causes agents to require less training data and reach higher asymptotic performance than if they had been trained on randomly sampled tasks \cite{PLR, ACCEL}. This exciting new research direction leads to several open problems. One problem is measuring an agent's generalisation ability over different regions of the target task distribution. This ability would allow us to intelligently sample portions of the target task distribution for the agent's future training. Furthermore, it remains unclear how to best sample these ``weak point" regions of the target task distribution such that the agent receives a meaningful learning signal. 

Lastly, a worthwhile research direction is to investigate formulations of objective functions directly adapted to the generalisation objective. For example, we could choose an objective that ensures the agent does not fail at any tasks in the target distribution instead of simply maximising expected returns across the target task distribution. It may not be possible to define this objective as a single real-valued function to maximise. Instead, it has been proposed to formulate the generalisation objective as optimising a performance distribution over the task space by Pareto optimality \cite{XLAND} or as a game played between the agent and a task generator \cite{PAIRED, Pinto2017RobustAdversarialRL}. However, the space of possible generalisation objectives remains largely unexplored, and some objectives may be more adapted than others to achieve specific types of generalisation.

\subsection{Causal Reinforcement Learning}
While most approaches proposed to address generalisation are centred around model-free RL, model-based RL methods have also shown success in solving many tasks \cite{modelbasedRLsurvey}. One way to leverage the benefits of learning such models in a structured way is to learn causal models of the environments. However, it is difficult to provide an RL agent with an accurate causal model in real-world scenarios, so the agent must learn the model through its interactions. Multi-agent systems provide a practical setting to learn causal models because the agents can access observational data from observing other agents and interventional data from performing their own actions. Together, these can provide the required dataset for causal structure learning. Furthermore, access to the causal model for decision-making is beneficial in multi-agent systems because it allows predictions to be adjusted for unobserved confounders to handle the unobserved factors affecting the decisions of other agents.

In addition to learning the causal model directly, causality techniques could be used to improve the representations learned by RL agents in image-based environments. Disentangled representation learning aims to separate distinct, informative factors of variation in an image to identify the ground-truth factors that generated the image. Access to a disentangled representation could improve policy learning and generalisation to unseen tasks~\cite{dunion2022ted}. However, it remains an open problem to learn such a representation in an RL setting.

\subsection{Open Challenges in Ad Hoc Teamwork}
Although our group has made progress towards solving the full ad hoc teamwork challenge, particularly in the case of open team settings, most of the work in the literature addresses the problem under different assumptions that may not hold in real-world applications. One of such assumptions is that the learner has full observability of the environment, which is not a realistic assumption for most robotics cases. In~\cite{Carlucho2022UnderwaterAHT}, we discussed how ad hoc teamwork agents can aid in marine applications, particularly in search and rescue operations. We evaluated current challenges that need to be addressed in order to achieve such a system, and we discussed future research directions. 
We argued that developing ad hoc agents that are able to communicate is of critical importance~\cite{macke2021expected}, but that these agents must be able to do so under realistic restrictions imposed by the available communication channels.

Another aspect of the ad hoc challenge that is not usually addressed is cases in which the learner needs to work with teammates that are able to learn or adapt their behaviour over time. Changes in the policies or in the behaviour of teammates can create uncertainties that can make learning difficult for the ad hoc agent.
This in turn highlights another challenge faced by ad hoc teamwork methods, that of teammate generation. During the learning stages, to achieve general policies that can be effective against a large number of teammates, the ad hoc agent needs to encounter and interact with teammates with different policies. Generally, teammates' policies are developed manually; however, as the tasks to solve become more and more complex, generating teammates becomes a more arduous task \cite{Jacob_Devlin_Hofmann_2020}. Currently, we are exploring different alternatives for the automatic generation of teammates that might help the learner to generalise more robustly to a wide set of possible teammates~\cite{Rahman2022TG}. 

Recently, at the 2022 IJCAI Workshop on ad hoc teamwork, we proposed a novel problem related to ad hoc teamwork called \textit{few-shot teamwork}~\cite{fosong2022fewshot}. This is a special case in which teams of agents, which have been trained independently to each solve a different task, are then combined together into a new team that has to learn and adapt to solve an unseen but related task.
The capability to adapt to new tasks in the manner could also help to accelerate MARL by making it possible to decompose a complex task into simpler sub-tasks with smaller teams, and later combining agents skilled at these sub-tasks and training on the complex task.

\subsection{Autonomous Driving}

While our goal recognition method for autonomous vehicle goal recognition, GRIT, has shown some advantages with respect to other methods, there are still some open problems. 
On main concern is that GRIT assumes full observability of the scene. However, in autonomous driving, there is often missing information due to occlusions. Therefore, GRIT requires specialised decision trees to be trained in advance for each possible goal location in fixed scenarios. In recent and ongoing work we addressed these issues using a new goal recognition method named \textit{Goal Recognition with Interpretable trees under Occlusion} (OGRIT)~\cite{brewitt2022verifiable} which builds on the GRIT method. OGRIT is designed to handle occlusions and generalise to unseen scenarios, while still being fast, accurate, interpretable and verifiable. Another open problem is improving the accuracy of interpretable and verifiable goal recognition methods. In some scenarios, a deep learning baseline was found to achieve slightly higher accuracy than GRIT~\cite{brewitt2021grit}, so there is still room to improve the accuracy of methods such as GRIT.

Another important question related to autonomous driving is whether we can generate intuitive dialogue-oriented explanations automatically in a way that builds \textit{transparency} and \textit{trust} in our system while making sure we match the passengers' expectations and correctly develop their understanding of our system.
IGP2 is particularly well-suited for building trust in autonomous vehicles, as it was designed from the start with interpretability in mind.
Using this inherent interpretability, we were able to extract intuitive explanations for the actions of autonomous vehicles powered by IGP2 \citep{albrecht2020igp2}.
Our next step is to build an automatic explanation generation system on top of IGP2 that can create such intuitive explanations automatically in response to natural language queries by human users.
In a paper which came runner-up for the best paper award at the 2022 IJCAI Workshop on Artificial Intelligence for Autonomous Driving we proposed an initial prototype system for such an explanation generation system~\cite{gyevnar2022humancentric}. 

While these issues relate to autonomous agents driving in urban environments, we are also interested in cases in which agents need to navigate through outdoor unstructured environments. However, outdoor environments have a high degree of variation, for example, changes in seasons, weather conditions, and the general position of objects. Therefore, in order to develop learning agents for navigating these types of environments a large amount of data is needed. In order to accelerate the development of such types of agents we recently introduced MIDGARD~\cite{vecchio2022midgard}, a simulation environment specifically designed for navigation in cluttered outdoor environments. MIDGARD utilises the Unreal Engine to achieve photorealistic scenarios. Additionally, it provides different scenes, such as Forests and Meadows, and supports the procedural generation of objects. We hope that the introduction of this simulator will help further the research in outdoor navigation.

\section{Conclusion}
\label{sec:conclusion}

The development of autonomous agents which can interact effectively with other agents to accomplish a given task is a core area of research in artificial intelligence and machine learning. Towards this goal, the Autonomous Agents Research Group develops novel machine learning algorithms for autonomous systems control, with a specific focus on deep reinforcement learning and multi-agent reinforcement learning. The group's research has been at the forefront of these areas, including scalable learning of coordinated agent policies and inter-agent communication; reasoning about the behaviours, goals, and composition of other agents from limited observations; and sample-efficient learning based on intrinsic motivation, curriculum learning, causal inference, and representation learning. This article provided an overview of the ongoing research portfolio of the group and discussed open problems for future research directions.

\section*{Acknowledgements}

Research in the Autonomous Agents Research Group has been funded by: UK Research and Innovation (UKRI), UK Engineering and Physical Sciences Research Council (EPSRC), Alan Turing Institute (ATI), Royal Society, Royal Academy of Engineering (RAEng), Defense Advanced Research Projects Agency (DARPA), US Office of Naval Research (ONR), and industry sponsors Google, Five AI, and Dematic/KION.

%

\bibliographystyle{ios1}           

\bibliography{bibliography}        

%

\end{document}